\documentclass[jkps,preprint,fleqn,showpacs,showkeys]{revtex4}
\usepackage{graphicx}
\usepackage{amssymb}
\usepackage{amsmath}
\usepackage{bm}

\usepackage{amsfonts,amssymb,amsmath}
\usepackage{graphicx}

\usepackage{graphics,hyperref}
\usepackage{amssymb}
\usepackage{amsmath}
\usepackage{color}
\usepackage{bm}

\begin{document}
\setcounter{page}{0}
\title[]{A brief review of Scott measure as a multipartite entanglement criterion}
\author{Saeed \surname{Haddadi}}
\email{haddadi@physicist.net}

\affiliation{Department of Physics, Payame Noor University, P.O. Box 19395-3697, Tehran, Iran.}

\begin{abstract}
In recent decades, various multipartite entanglement measures have been proposed by many researchers, with different characteristics. Meanwhile, Scott studied various interesting aspects of multipartite entanglement measures and he has defined a class of related multipartite entanglement measures in an obvious manner. Recently, Jafarpour \textit{et al.} (\href{https://doi.org/10.1142/S0219749915500471}{Int. J. Quantum Inform. \textbf{13}, 1550047 2015}) have calculated the entanglement quantity of two-dimensional 5-site spin system by means of Scott measure. As it was presented in their calculation $Q_{2}>Q_{3}$. In this paper, we would like to point out that in the 5-qudit system, $Q_{2}$ does not provide a stronger entanglement than $Q_{3}$, while there is a simple relation between them.
\end{abstract}



\maketitle

\section*{}

In 2004, Scott \cite{Scott03} has investigated the average bipartite entanglement, over all possible divisions of a multipartite system, as a useful multipartite entanglement measure \cite{Haddadi}. The Scott measure (or generalized Meyer-Wallach measure) for $N$-qudit system ($d$-dimensional) is defined as follows \cite{Love02,Haddadi,Scott03}
\begin{equation}
\label{E1}
Q_{m}(|\psi\rangle)\equiv\frac{d^{m}}{d^{m}-1}\Big(1-\frac{m!(N-m)!}{N!}\sum_{|S|=m}\mathrm{Tr}[\rho_{S}^{2}]\Big), \quad m=1,\cdots,\lfloor N/2 \rfloor
\end{equation}
where $S\subset \{1,\cdots, N\} $ and $\rho_{S}$ = $\mathrm{Tr}_{\acute{S}}(|\psi\rangle\langle\psi|)$ is the reduced density matrix for qudits $S$ after tracing out the rest ($d$=2 for two-dimensional spin system). The $Q_{m}$ quantities correspond to the average entanglement between subsystems which consists $m$ qudits and the remaining $N-m$ qudits \cite{Borras04}.

Recently, Jafarpour \textit{et al.} \cite{Jafarpour01} based on Scott measure have calculated the entanglement quantity in two-dimensional 5-site spin system. They have calculated $Q_{2}$ and $Q_{3}$ in Table 1 and Table 2, however, there is a relation between them fixed by the Schmidt decomposition \cite{Schmidt05}; namely the eigenvalues of the reduced density matrix of qudits 12, will be the same as the eigenvalues of the reduced density matrix of qudits 345. In other words, $\mathrm{Tr}[\rho_{ij}^{2}]=\mathrm{Tr}[\rho_{klm}^{2}]$ with $i,j,k,l,m \in \{1,\cdots,5\}$ for all 5-qudit pure states \cite{Fei06}, which this implies a simple relation between $Q_{2}$ and $Q_{3}$. We emphasize that in a 5-qudit system, $Q_{2}$ and $Q_{3}$ both capture entanglement in partition 2 qudits vs 3 qudits. We have also shown in Ref \cite{Haddadi07} that for a 5-qudit system, $Q_{2}$ and $Q_{3}$ both refer to 3-2 partitions of the same state. Therefore, Jafarpour \textit{et al.} \cite{Jafarpour01} could not say that $Q_{2}>Q_{3}$ for all states in a two-dimensional 5-site spin system.

In conclusion, the above analysis shows that the authors \cite{Jafarpour01} did not use the Scott measure properly. Since $Q_{2}$ and $Q_{3}$ are proportional to each other by a numerical factor (as seen from (\ref{E1})). In fact, for the 5-qudit system, $Q_{1}$ refers to 1-4 partitions and $Q_{2}$ or $Q_{3}$ both refer to 3-2 partitions of the same state, which is a very significant point. Hence, we conclude that $Q_{2}$ does not provide a stronger entanglement than $Q_{3}$ in the 5-qudit system.

\end{document}